\documentclass[reprint,aps,groupedaddress,showpacs]{revtex4-1}
\usepackage{graphicx}
\usepackage{amsmath,amssymb,lmodern}
\usepackage{amsfonts}
\usepackage{amssymb}
\usepackage{bm}
\usepackage{color}

\begin{document}

\title{Plasmonic detection of the parity anomaly in a two-dimensional Chern insulator}
\author{M. N. Chen}
\email{mnchen@hdu.edu.cn}
\author{Yu Zhou}
\email{yzhou@hdu.edu.cn}
\affiliation{$^1$School of Science, Hangzhou Dianzi University, Hangzhou, 310018, China}

\begin{abstract}
  In this paper, we present an analytic study on the surface plasmon polaritons in two-dimensional parity anomaly Chern insulators. The two-dimensional conductivity derived from the BHZ model are antisymmetric, based on which two surface plasmon modes each contains two branches of dispersions have been found. In the absence of parity anomaly, the Hall conductivities with positive and negative Dirac mass terms differ by a sign; two branches of each surface plasmon mode are exactly degenerate. However, the parity anomaly can lift such degeneracy and lead to significant modifications of these dispersion curves or even the occurrence of an extra branch of surface plasmons under particular condition. Our investigations pave a possible way for the detection of the parity anomaly in a two-dimensional Chern insulator via plasmonic responses.
\end{abstract}

\maketitle



\section{\label{section1}INTRODUCTION}
Topological materials have attracted much attention
in both theoretical and experimental aspects
in recent years. As a typical class of topological materials,
topological insulators (TIs) have exotic
metallic surface (boundary) states protected by time-reversal symmetry,
whose topological charge is identified as the $\mathbb{Z}_2$ invariant~\cite{TI1}.
Unlike TIs, the Chern insulators (CIs), or named as quantum anomalous Hall insulators,
break the time-reversal symmetry, which belongs to the $\mathbb{Z}$-topological classification
and the corresponding topological charge is the first Chern number.
A spin-conserved TI may be viewed as two copies of CIs carrying opposite spin polarizations
and counter propagating edge states, respectively~\cite{TI2,TI3}.

Both TIs and CIs have nontrivial responses to external electromagnetic fields.
For TIs, we have $j^{\mathrm{s}}_\mu=\sigma^{\mathrm{s}}_{xy}\epsilon_{\mu\nu\tau}
\partial^\nu\Omega^\tau$ with $\sigma^{\mathrm{s}}_{xy}$ the spin-Hall
conductivity and $\Omega$ being a pure gauge~\cite{BHZ,BHZ1,res1,res2}; for CIs,
it has form $j_\mu=\sigma_{xy}\epsilon_{\mu\nu\tau}A^\mu\partial^\nu A^\tau$ with
$\sigma_{xy}$ the Hall conductivity and $A^\mu$ the gauge fields. Therefore,
nontrivial electromagnetic responses originate from topologically-nontrivial
bulk band structures, which may lead to nontrivial collective excitations.

Surface plasmons are collective oscillations of free electrons coupled with light existing at the
metal-dielectric interface, of which the electric fields are tightly confined and decay exponentially away from the surface~\cite{SPPbook,JacksonBook}. The permittivities at two sides usually possess opposite signs, i.e. one is positive and the other is negative; otherwise, no dispersion relations can be found for these surface waves. After the discovery of graphene, researchers have realized that such one-atom thick material can support exceedingly strong surface plasmons that is detectable through, for example, scanning near-field infrared microscopy~\cite{GrapRev1,GrapRev2,GrapRev3,NatureGrap1,NatureGrap2,NSOM,NatureGrap3}. This can be understood that the conductivity of doped graphene is large enough to cause significant in-plane currents and charge oscillations under incident light pushing the corresponding Drude plasma frequency into the infrared region~\cite{G1,G2,G3,G4,G5,G6,G7,G8,G9,G10,G11,G12}. Although the optical conductivity of graphene is isotropic, its imaginary part can be positive or negative depending on the Fermi level as well as the photon energy. With a positive imaginary part, graphene resembles a thin metallic film supporting transverse magnetic- (TM) polarized surface plasmons; however, with a negative imaginary part, it is more like a thin dielectric film and the surface plasmons are transverse electric- (TE) polarized~\cite{G13}. Due to the isotropy, TM- and TE-polarized modes are decoupled and their dispersion relations can be found separately.

Anisotropy can lead to the coupling of these two polarizations. For example, in phosphorene the electron masses are largely different along the zigzag and armchair directions due to its puckered structure. After being doped with electrons, phosphorene can become metallic supporting surface plasmons~\cite{BP1,BP2,BP3,BP4}; the optical conductivities differ along the zigzag and armchair directions. The iso-frequency contour of the in-plane surface plasmons is in most cases elliptic. With proper electron doping, the conductivities along two directions can have opposite signs and the corresponding iso-frequency contour becomes a hyperbola. In this case, they are called hyperbolic surface plasmons~\cite{BP5}. In the calculation of dispersion relations, one must solve all the field components since the anisotropy usually mixes the two polarizations mentioned above.

As for the systems with nonzero Hall conductivities such as CIs, the off-diagonal terms of the conductivity tensor induce currents orthogonal to the applied electric fields, which immediately leads to the situation where all the field components are interrelated and should be all considered simultaneously during the calculation of the surface plasmons~\cite{TIs1,TIs2,TIs3,TIs4,TIs5,TIs6}. The dispersion relation strongly depends on the conductivity as well as the permittivity of the surrounding materials. Most 2D CIs are encapsulated with optically anisotropic dielectrics showing different permittivities parallel and perpendicular to the conductive surface. Such anisotropy can significantly modify the dispersion relations of the surface plasmons as well.

\begin{figure}
  \centering
  \includegraphics[width=3.0in]{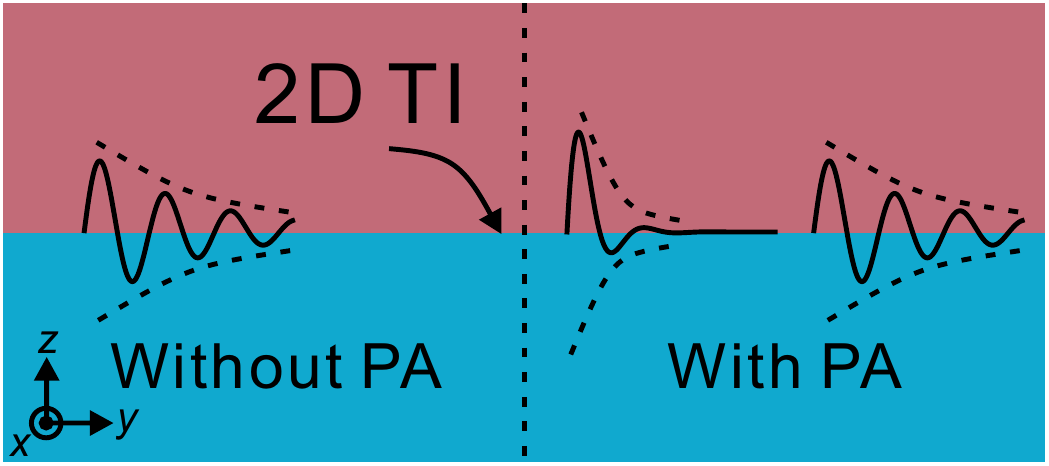}
  \caption{Schematic illustration of the detection of the parity anomaly in 2D CIs. An extra branch of the surface plasmons may occur under particular condition with parity anomaly. Left/right panel corresponds to the the case without/with parity anomaly.}
  \label{fig1}
\end{figure}

In this paper, we intend to reveal the connections between the parity anomaly in a two-dimensional Chern insulator and the dispersion relations of the surface plasmons. Given by the symmetry of the BHZ model, two surface plasmon modes have been found, each of which contains two branches of dispersion relations. Two modes are respectively characterized by $E_z=0$ and $H_z=0$;  the expressions of their dispersion relations indicate that two branches of each mode are exactly degenerate without parity anomaly. One arrives at the same dispersion curve with positive and negative Dirac mass terms. However, in the presence of the parity anomaly, such degeneracy would be lifted; the Hall conductivities in the topologically trivial and non-trivial situations no longer just differ by a sign and an extra branch of surface plasmons might be found under particular condition, as schematically shown in Fig.\ref{fig1}. Our investigations pave a possible way for the detection of the parity anomaly in a two-dimensional Chern insulator via plasmonic responses.

This paper has been organized as follows. In Sec.\ref{section2}, details regarding the BHZ model describing the two-dimensional Chern insulator and the calculations of the optical conductivities are given. The real and imaginary parts of both the longitudinal and Hall conductivities have been derived. In Sec.\ref{section3}, the expressions of the dispersion relations of the surface plasmons are presented based on the two dimensional conductivity tensor given by the BHZ model. Two modes each with two branches have been found. In Sec.\ref{section4}, the dispersion relations or equivalently the effective indices of the surface plasmons have been numerically calculated for all the cases. In Sec.\ref{section5}, a conclusion has been given.

\section{\label{section2}Model and Optical Conductivities}

Let us start from the minimal Hamiltonian for the Bernevig-Hughes-Zhang (BHZ) model,
which can be written as~\cite{BHZ,BHZ1}
\begin{equation}
\hat{H}=v\hbar(k_x\hat{\sigma}_x+k_y\hat{\sigma}_y)+m_{\mathbf{k}}\hat{\sigma}_z\ ,
\label{BHZ}
\end{equation}
where $v$ stands for the Fermi velocity,
$m_{\mathbf{k}}=mv^2-b\hbar^2k^2$ is the regularized Dirac mass term with $k^2=k^2_x+k^2_y$,
and $\{\hat{\sigma}_i\}$ are Pauli matrices with $i=x,y,z$.
The Hamiltonian (\ref{BHZ}) is usually used to describe the Chern insulator,
where we have assumed that the spin is fully polarized and thus the spin freedoms can be ignored.

After straightforward diagonalization,
one can obtain the eigenvalues:
\begin{equation}
\epsilon_\pm(k)=\pm\sqrt{v^2\hbar^2k^2+m^2_{\mathbf{k}}}\ ,
\end{equation}
and the corresponding eigenvectors are
\begin{equation}
|u_-\rangle=\begin{pmatrix}\sin\theta_k\mathrm{e}^{-\mathrm{i}\varphi_k}\\
-\cos\theta_k\end{pmatrix},\quad
|u_+\rangle=\begin{pmatrix}\cos\theta_k\mathrm{e}^{-\mathrm{i}\varphi_k}\\
\sin\theta_k\end{pmatrix},
\label{eigen1}
\end{equation}
where $\varphi_k=\arg(k_x+\mathrm{i}k_y)$ and $2\theta_k=\mathrm{arccot}(m_{\mathbf{k}}/v\hbar k)$.

\begin{figure*}
  \centering
  \includegraphics[width=4.8in]{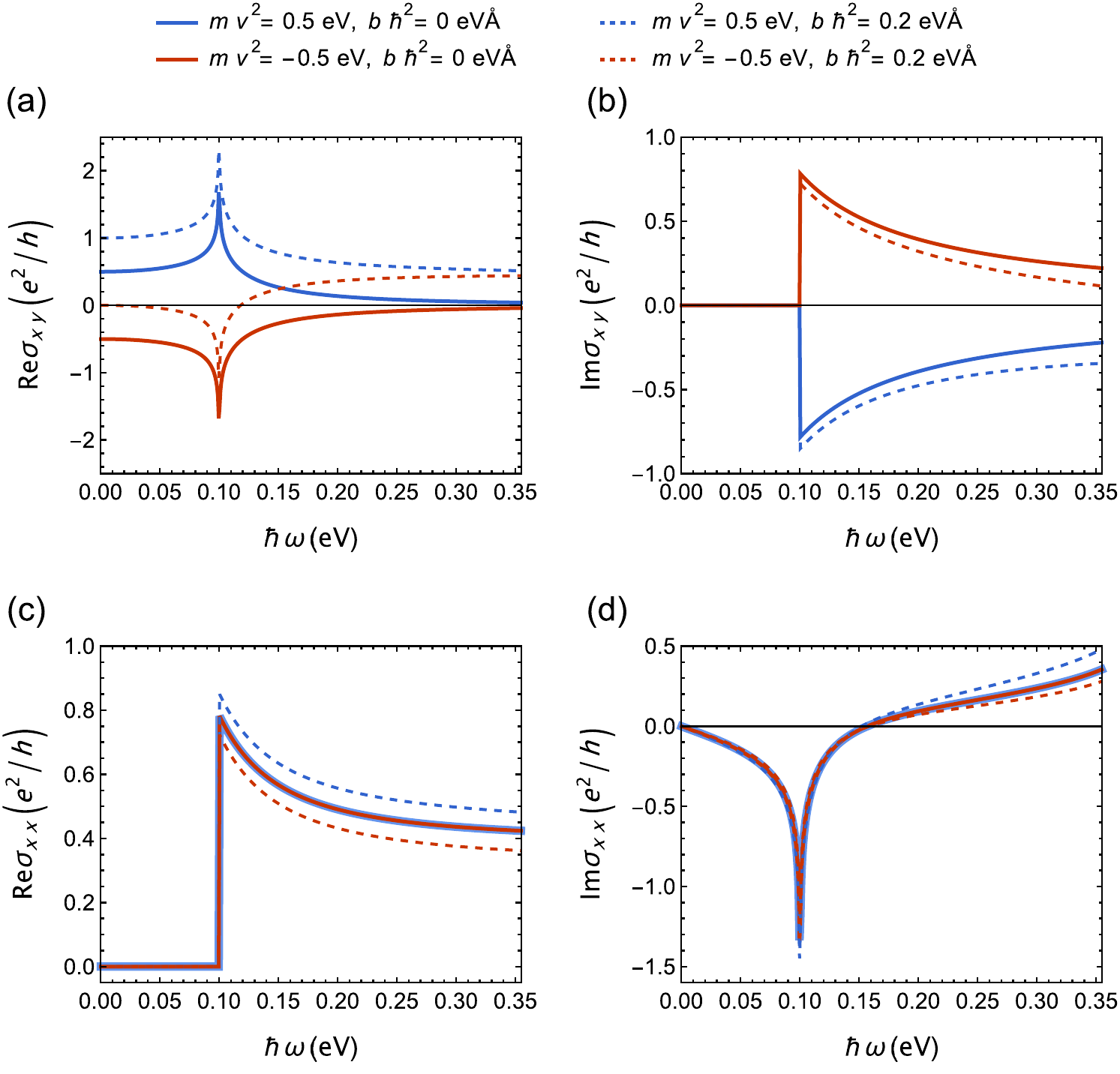}
  \caption{Plots of conductivities (a) $\mathrm{Re}\sigma_{xy}$, (b) $\mathrm{Im}\sigma_{xy}$, (c) $\mathrm{Re}\sigma_{xx}$, and (d) $\mathrm{Im}\sigma_{xx}$ as functions of the photon energy $\hbar\omega$ in units of $e^2/h$. Solid and dashed curves correspond to the cases without and with parity anomaly, respectively. The conductivities with positive and negative Dirac mass terms are plotted in blue and red, respectively. The cut-off energy is $\epsilon_{\mathrm{c}}=4|m|v^2$ in (d). Other parameters are $\hbar v=0.5$eV, $\hbar^2b=0.2$eV$\cdot$\AA$^2$.}
  \label{fig2}
\end{figure*}

\begin{figure}
  \centering
  \includegraphics[width=3.2in]{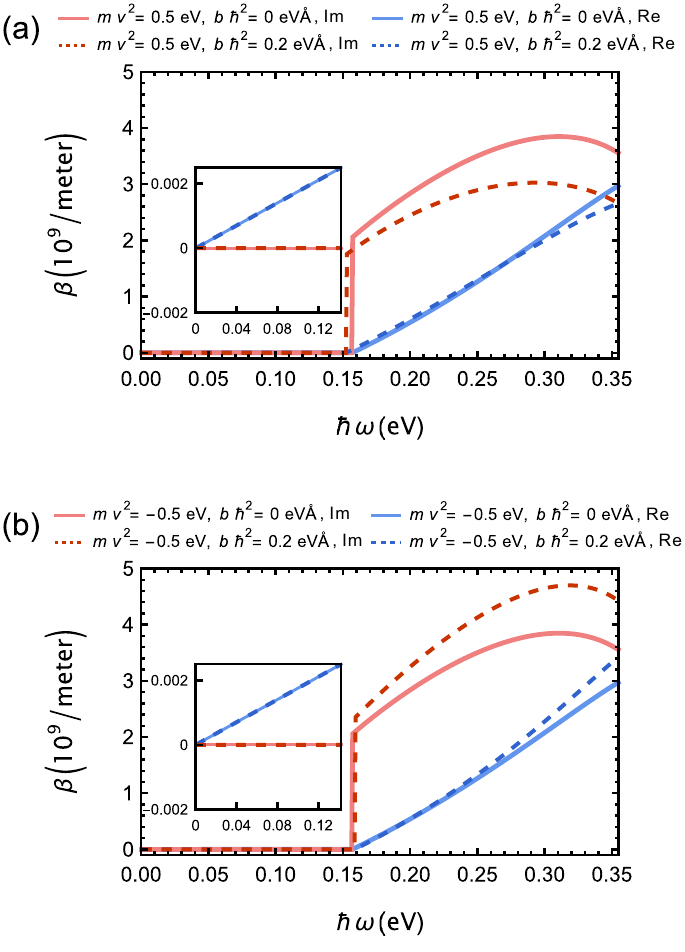}
  \caption{Propagation constants of the surface plasmon mode with $H_z=0$ for the situations with (a) $mv^{2}=0.05$ and (b) $mv^{2}=-0.05$. Insets are the enlarge plots in the low photon energy ranges. The solid and imaginary parts are plotted in blue and red, respectively. Propagation constants correspond to the case with/without parity anomaly are plotted with dashed/solid curves.}
  \label{fig3}
\end{figure}

\begin{figure*}
  \centering
  \includegraphics[width=4.8in]{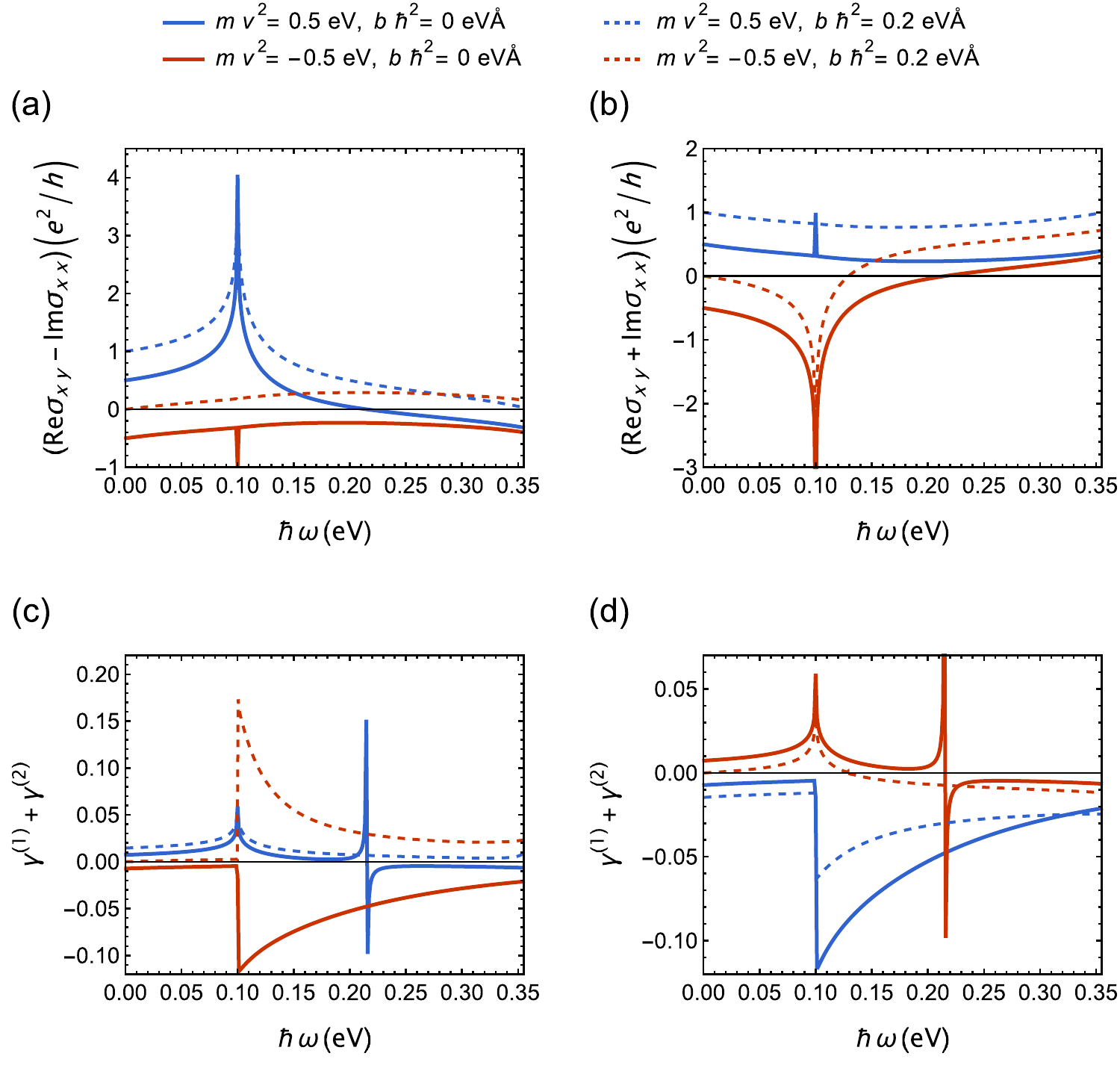}
  \caption{(a) $\mathrm{Re}\sigma_{xy}-\mathrm{Im}\sigma_{xx}$ and (b) $\mathrm{Re}\sigma_{xy}+\mathrm{Im}\sigma_{xx}$ as functions of the photon energy. Two branches of dispersion relations given by Eq.(\ref{set1}) and Eq.(\ref{set3}) require $\mathrm{Re}\sigma_{xy}-\mathrm{Im}\sigma_{xx}>0$ and $\mathrm{Re}\sigma_{xy}+\mathrm{Im}\sigma_{xx}<0$, respectively. (c) and (d) are $\gamma^{(1)}+\gamma^{(2)}$ corresponding to Eq.(\ref{set1}) and Eq.(\ref{set3}) as functions of the photon energy, respectively. It must be satisfied that $\gamma^{(1)}+\gamma^{(2)}>0$.}
  \label{fig4}
\end{figure*}

\begin{figure*}
  \centering
  \includegraphics[width=4.8in]{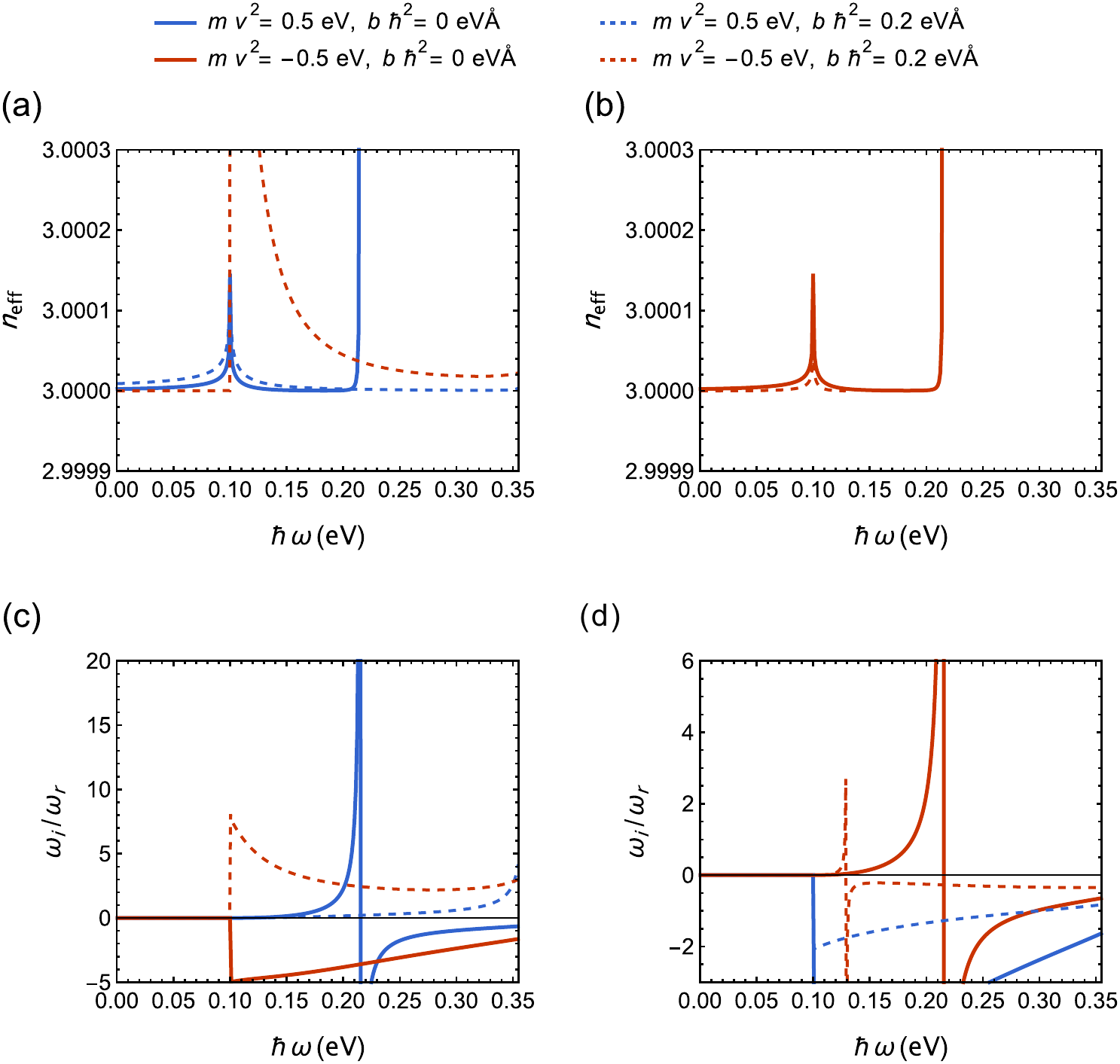}
  \caption{Effective indices as functions of the photon energy for the two branches of surface plasmons given by (a) Eq.(\ref{set1}) and (b) Eq.(\ref{set3}).
  The ratios $\omega_{i}/\omega_{r}$ of the two branches of surface plasmons given by (c) Eq.(\ref{set1}) and (d) Eq.(\ref{set3}).}
  \label{fig5}
\end{figure*}

The optical conductivity tensor can be obtained by the standard Kubo formula~\cite{Rammer}
for the $d$-dimensional system,
\begin{equation}
\sigma_{ij}(\omega)=-\frac{\mathrm{i}}{L^d}\frac{e^2}{\hbar}
\sum_{n,m}\frac{[n_{\mathrm{F}}(\epsilon_n)-n_{\mathrm{F}}(\epsilon_m)]v^i_{nm}v^j_{mn}}
{(\epsilon_n-\epsilon_m)(\epsilon_n-\epsilon_m+\hbar\omega)}
\label{kubo}
\end{equation}
where $v^i_{nm}=\langle u_n|\frac{\partial\hat{H}}{\partial k_i}|u_m\rangle$
is the velocity operator in the $i$-th direction with $i=1,2,\ldots,d$.
The indices $m,n=+,-$ represent the conduction band and valence band, respectively,
$n_{\mathrm{F}}(\epsilon_n)=1/\big(1+\mathrm{e}^{\beta(\epsilon_n-\mu)}\big)$
is the Fermi-Dirac distribution function with $\mu$ being the chemical potential,
$\beta=1/(k_{\mathrm{B}}T)$ with $k_{\mathrm{B}}$ being the Boltzmann constant and
$T$ is the tmeperature, and $L$ is the length of the system.
For the model we have considered in this work, $d=2$.
In what follows, we will include the impurity scattering processes,
therefore, the frequency $\omega$
should be replaced by $\omega+\mathrm{i}/2\tau$ with $\tau$ the
elastic scattering time.
Substituting Eq.(\ref{eigen1}) into Eq.(\ref{kubo})
and after doing some algebras, the expression for the optical conductivity becomes
\begin{widetext}
\begin{equation}
\begin{split}
\sigma_{ij}(\omega)&=-\mathrm{i}e^2\hbar\int\frac{\mathrm{d}^2\mathbf{k}}
{(2\pi)^2}\frac{n_{\mathrm{F}}(\epsilon_+)-n_{\mathrm{F}}(-\epsilon_+)}{2\epsilon_+}\\&\quad\times
\bigg[v^i_{+-}v^j_{-+}\frac{\hbar\omega-2\epsilon_+}{(\hbar\omega-2\epsilon_+)^2+\hbar^2/4\tau^2}
-\mathrm{i}\pi v^i_{+-}v^j_{-+}\delta(\hbar\omega-2\epsilon_+)
+v^i_{-+}v^j_{+-}\frac{1}{2\epsilon_++\hbar\omega}\bigg]\ .
\end{split}
\label{kubo1}
\end{equation}
Using Eq.(\ref{kubo1}), we can obtain the real part for the
longitudinal conductivity $\sigma_{xx}$ at zero temperature as (in units of $e^2/h$)
\begin{equation}
\begin{split}
\mathrm{Re}\,\sigma_{xx}(\omega)=\frac{\pi v^2}{8}
\frac{1}{v^2-2b(mv^2-b\hbar^2k^2)}
\bigg[1+\frac{4(mv^2+b\hbar^2k^2)^2}{(\hbar\omega)^2}\bigg]\Theta(\hbar\omega-2|m|v^2)
\end{split}
\end{equation}
\end{widetext}
where $k^2$ is solved via the equation
\begin{equation}
\frac{(\hbar\omega)^2}{4}=v^2\hbar^2k^2+(mv^2-b\hbar^2k^2)^2\ ,
\end{equation}
the Fermi energy $\epsilon_{\mathrm{F}}$ is set to zero
(i.e. stays in the band gap and hence the intraband contribution to the conductivity is zero),
and the imaginary part of $\sigma_{xx}$ is given by
\begin{widetext}
\begin{equation}
\mathrm{Im}\,\sigma_{xx}(\omega)=\frac{v^2}{4}\int\mathrm{d}\epsilon_+
\frac{1}{\sqrt{(1-4bm)v^4+4b^2\epsilon^2_+}}
\bigg[1+\frac{(mv^2+b\hbar^2 k^2)^2}{\epsilon^2_+}\bigg]
\Bigg[\frac{\hbar\omega-2\epsilon_+}{(\hbar\omega-2\epsilon_+)^2
+\frac{\hbar^2}{4\tau^2}}+\frac{1}{2\epsilon_++\hbar\omega}\Bigg]
\end{equation}
The real part of the Hall conductivity $\sigma_{xy}$ can be derived as
\begin{equation}
\mathrm{Re}\,\sigma_{xy}(\omega)=\frac{v^2}{4b\hbar\omega\xi}
\sum_{s=+,-}(1-4bm+s\xi)
\mathrm{arccoth}\Bigg[\frac{2\frac{(b\hbar)^2k\omega}{v^3}\sqrt{1+\big(\frac{mv}{\hbar k}
-\frac{b\hbar k}{v}\big)^2}}{s(1-4bm)+(1-2bm+\frac{2b^2k^2\hbar^2}{v^2})\xi}\Bigg]\Bigg|^\infty_0
\label{hall1}
\end{equation}
\end{widetext}
and the imaginary part of $\sigma_{xy}$
\begin{equation}
\mathrm{Im}\,\sigma_{xy}(\omega)=-\frac{\pi v^2}{2\hbar\omega}
\frac{mv^2+bk^2\hbar^2}{v^2-2b(mv^2-b\hbar^2k^2)}\Theta(\hbar\omega-2|m|v^2)
\end{equation}
where $\xi=\sqrt{1-4bm+(b\hbar\omega/v^2)^2}$.
Using the relation
$\mathrm{arccoth}(x)=[\ln((x+1)/x)-\ln((x-1)/x)]/2$,
Eq.(\ref{hall1}) becomes
\begin{equation}
\begin{split}
\mathrm{Re}\,\sigma_{xy}(\omega)=&\frac{v^2}{8\xi b\hbar\omega}\times\\&\Bigg[2(1-4bm)
\ln\bigg|\frac{b\hbar\omega/v^2+\xi}{b\hbar\omega/v^2+\xi}\bigg|-\sum_{s=+,-}g_{s}(\omega)\Bigg]\ ,
\end{split}
\end{equation}
where we have defined
\begin{equation}
\begin{split}
g_{s}(\omega)=&\big(1-4bm+s\xi\big)\times\\&\ln
\bigg|\frac{2b^2|m|\hbar\omega/v^2+(1-4bm)s+\xi(1-2bm)}{2b^2|m|\hbar\omega/v^2-(1-4bm)s-\xi(1-2bm)}\bigg|\ .
\end{split}
\end{equation}
In the DC limit ($\omega\rightarrow0$), the real part of $\sigma_{xy}$ becomes
\begin{equation}
\mathrm{Re}\,\sigma_{xy}
=\frac{e^2}{h}C\ ,
\end{equation}
where $C=[\mathrm{sgn}(m)+\mathrm{sgn}(b)]/2$ is the first Chern number,
which characterizes the topological properties of the Hamiltonian (\ref{BHZ}), as expected.
Equation (\ref{hall1}) agrees with the results obtained in Ref.\cite{Shen1}.
All of the conductivities as functions of the photon energy $\hbar\omega$ are shown in Fig.\ref{fig2}.
One can find that
there exist peaks at $\hbar\omega=2mv^2$, which are due to the Rabi resonance.
One may also see from Figs.\ref{fig2}(c) and \ref{fig2}(d)
that the sign of $m$ does not affect the longitudinal conductivities qualitatively,
while it can significantly modify the Hall conductivities.
This can be understood by fact that the topological
nature for the Hall conductivity
depends on the relative sign between the values of $m$ and $b$,
which also determines the values of the Chern number $C$.

In the absence of the parity anomaly term ($b=0$), the conductivities
can be reduced to
\begin{equation}
\mathrm{Re}\,\sigma_{xy}(\omega)=\frac{e^2}{h}\frac{mv^2}{2\hbar\omega}
\ln\bigg|\frac{2v^2|m|+\hbar\omega}{2v^2|m|-\hbar\omega}\bigg|\ ,
\end{equation}
\begin{equation}
\mathrm{Im}\,\sigma_{xy}(\omega)=-\frac{e^2}{h}\frac{\pi}{2}\frac{mv^2}
{\hbar\omega}\Theta(\hbar\omega-2|m|v^2)\ ,
\end{equation}
\begin{equation}
\mathrm{Re}\,\sigma_{xx}=\frac{e^2}{h}\frac{\pi}{8}
\bigg(1+\frac{4m^2v^4}{(\hbar\omega)^2}\bigg)\Theta(\hbar\omega-2|m|v^2)
\end{equation}
and
\begin{equation}
\begin{split}
\mathrm{Im}\,\sigma_{xx}(\omega)&=
\frac{e^2}{h}\frac{1}{4}\int\mathrm{d}\epsilon_+\bigg(1+\frac{m^2v^4}{\epsilon^2_+}\bigg)
\\&\quad\times\Bigg[\frac{\hbar\omega-2\epsilon_+}{(\hbar\omega-2\epsilon_+)^2
+\frac{\hbar^2}{4\tau^2}}+\frac{1}{2\epsilon_++\hbar\omega}\Bigg]\ .
\end{split}
\end{equation}
In fact, from the point view of the topological field theory,
the Hamiltonian (\ref{BHZ}) is reduced to a
(2+1)-dimensional massive Dirac one when $b=0$, whose mass term $mv^2\hat{\sigma}_z$
plays an important role in Chern-Simons field Lagrangian~\cite{1996Aspects}
\begin{equation}
\mathcal{L}_{\mathrm{CS}}=\frac{\mathrm{sgn}(m)}{2}\int\mathrm{d}^2x\mathrm{d}t\,
\epsilon^{\mu\nu\tau}A_\mu\partial_\nu A_\tau\ ,
\label{CS1}
\end{equation}
where $A_\mu$ is the gauge field with the space-time indices $\mu=t,x,y$.
The presence of the regulating term $b\hbar^2k^2$ results in an additional term $\mathrm{sgn}(b)/2$
to the coefficient of the Chern-Simons term (\ref{CS1}), which leads to the parity anomaly
and then Eq.(\ref{CS1}) becomes $\mathcal{L}_{\mathrm{CS}}=C\int\mathrm{d}^2x\mathrm{d}t\,
\epsilon^{\mu\nu\tau}A_\mu\partial_\nu A_\tau$.
In the following sections, we will consider the surface plasmonic responses
in both $C=1$ and $C=0$ cases, with and without parity anomaly.

\section{Surface plasmons}\label{section3}
The BHZ model gives us a two dimensional conductivity tensor
as shown above which should support surface plasmons.
Surface plasmons are coupled states of light and collective electron oscillations;
one should work with Maxwell's
equations at each side of the conductive layer and at same time consider the current density within the layer given by the conductivity tensor.
The dispersion relations of the surface plasmons can be derived based on the following two boundary conditions:
(i)  the tangential electric fields are continuous across the 2D CIs;
(ii) the current densities satisfy Ampere's
law which causes discontinuity of the tangential magnetic fields at two sides.
Due to the existence of the Hall conductivity, one cannot separate the surface plasmons
into transverse electric (TE) and transverse magnetic (TM) polarized modes;
in fact, they are coupled through $\sigma_{xy}$.
The Hall term has nothing to do with the ohmic losses but can seriously
modify the dispersion relations of the surface plasmons as shown below.
Considering the fact that the surrounding dielectrics at two sides are often composed of layered materials,
anisotropy is allowed where the in-plane and our-of-plane permittivities are respectively denoted as
$\varepsilon_{\mathrm{in}}$ and $\varepsilon_{\mathrm{out}}$.
The wave number $\mathbf{k}=(k_x,k_y,k_z)$ in the Cartesian coordinates can be separated into an in-plane part
$\boldsymbol{\beta}=(k_x,k_y)$ and a out-of-plane part $k_z$.
Since we are interested in the surface waves, it is common to set $k_z=\mathrm{i}\gamma$ with $\gamma$ being real.
In the surrounding dielectrics, one can derive the following expression from Maxwell's
equations considering the rotation symmetry of system implied by the BHZ model:
\begin{equation}
[(\gamma^2+k^2_0\varepsilon_{\mathrm{in}})(\beta^2-k^2_0\varepsilon_{\mathrm{out}})-\beta^2\gamma^2]
(\beta^2-\gamma^2-k^2_0\varepsilon_{\mathrm{out}})=0\ ,
\end{equation}
where $\beta=|\boldsymbol{\beta}|$ is the magnitude of the in-plane wave number,
$k_0=\omega/c$ is the wave number in vacuum with $\omega$
being the angular frequency. The above expression leads to two modes with either
$\beta^2-\gamma^2-k^2_0\varepsilon_{\mathrm{out}}=0$ or
$(\gamma^2+k^2_0\varepsilon_{\mathrm{in}})(\beta^2-\gamma^2-k^2_0\varepsilon_{\mathrm{out}})=\beta^2\gamma^2$.
Assuming $\gamma>0$, we have $\gamma=\sqrt{\beta^2-k^2_0\varepsilon_{\mathrm{out}}}$ or
$\gamma=\sqrt{(\beta^2/\varepsilon_{\mathrm{out}}-k^2_0)\varepsilon_{\mathrm{in}}}$
depending on the mode we are considering. The first expression clearly shows that
$\varepsilon_{\mathrm{out}}$ is not involved, thus $E_z=0$.
Further calculations of the polarization indicate that $\boldsymbol{\beta}\cdot\mathbf{E}=0$.
The second expression shows that $E_x$, $E_y$ and $E_z$ are all involved,
while further calculations indicate that $H_z=0$ and $\boldsymbol{\beta}\cdot\mathbf{H}=0$.
In searching the surface plasmons, we have chosen to solve $E_x$ and $E_y$.

Firstly, it is assumed that $E_z\neq0$, The electric fields are tightly confined near the conductive surface,
thus they are proportional to $\mathrm{e}^{\mathrm{i}k_xx}\mathrm{e}^{\mathrm{i}k_yy}\mathrm{e}^{-\gamma z}$,
and $E_z=\mathrm{i}\boldsymbol{\beta}\cdot\mathbf{E}\varepsilon_{\mathrm{in}}/\gamma\varepsilon_{\mathrm{out}}$.
Consequantly, $H_x$ and $H_y$ are given by
\begin{equation}
\begin{split}
-\frac{\varepsilon_{\mathrm{in}}}{\varepsilon_{\mathrm{out}}}\bigg(\frac{k_xk_y}{\gamma}E_x+\frac{k^2_y}{\gamma}E_y\bigg)
+\gamma E_y&=\mathrm{i}\omega\mu_0H_x\\
-\gamma E_x+\frac{\varepsilon_{\mathrm{in}}}{\varepsilon_{\mathrm{out}}}\bigg(\frac{k^2_x}{\gamma}E_x+\frac{k_xk_y}{\gamma}E_y\bigg)
&=\mathrm{i}\omega\mu_0H_y
\end{split}
\label{Hequation}
\end{equation}
where $\mu_0$ is the permeability of vacuum.
Based on the boundary conditions mentioned above, the equations regarding
$E_x$ and $E_y$ can be derived which finally leads to the following expression
for the dispersion relation of this surface plasmon mode
\begin{equation}
\begin{split}
&\big(\gamma^{(1)}+\gamma^{(2)}-k^2_x\Gamma-\mathrm{i}\omega\mu_0\sigma_{xx}\big)
\big(\gamma^{(1)}+\gamma^{(2)}-k^2_y\Gamma-\mathrm{i}\omega\mu_0\sigma_{yy}\big)
\\&=\big(k_xk_y\Gamma+\mathrm{i}\omega\mu_0\sigma_{xy}\big)
\big(k_xk_y\Gamma+\mathrm{i}\omega\mu_0\sigma_{yx}\big)\ ,
\end{split}
\label{dispersion1}
\end{equation}
where $\Gamma=\varepsilon^{(1)}_{\mathrm{in}}/(\gamma^{(1)}\varepsilon^{(1)}_{\mathrm{out}})
+\varepsilon^{(2)}_{\mathrm{in}}/(\gamma^{(2)}\varepsilon^{(2)}_{\mathrm{out}})$. $\gamma^{(i)}=\sqrt{\beta^2-k^2_0\varepsilon^{(i)}_{\mathrm{in}}}$, where the superscripts $i=1,2$ denote the space above and below the conductive layer, respectively.
The expression of the dispersion relation given above can be simply written as
\begin{equation}
\begin{split}
&\big(\gamma^{(1)}+\gamma^{(2)}-\mathrm{i}\omega\mu_0\sigma_{xx}\big)^2
-(\omega\mu_0\sigma_{xy})^2\\&=
\big(\gamma^{(1)}+\gamma^{(2)}-\mathrm{i}\omega\mu_0\sigma_{xx}\big)\beta^2\Gamma
\end{split}
\end{equation}
Solving $\beta$ in the complex plane with $\mathrm{Re}\,\beta\geq0$ and $\mathrm{Im}\,\beta\geq0$
at each frequency one can find the dispersion curves of the surface plasmons. Below the gap,
$\sigma_{xx}$ is purely imaginary and $\sigma_{xy}$ is purely real, hence
one only needs to solve $\beta$ on the real axis.

The above equation contains two branches of dispersion relations, which can be written as
\begin{widetext}
\begin{equation}
\frac{\omega}{c}=\frac{Z_0(\mathrm{i}\sigma_{xx})[2(\gamma^{(1)}+\gamma^{(2)})-\beta^2\Gamma]\pm
\sqrt{Z^2_0\sigma^2_{xy}[2(\gamma^{(1)}+\gamma^{(2)})-\beta^2\Gamma]^2+Z^2_0\beta^4\Gamma^2
[(\mathrm{i}\sigma_{xx})^2-\sigma^2_{xy}]}}
{2Z^2_0[(\mathrm{i}\sigma_{xx})^2-\sigma^2_{xy}]}
\label{equation21}
\end{equation}
\end{widetext}
where $Z_0$ is the vacuum impedance. Equation (\ref{equation21}) is one of our main results concerning the dispersion relations of the surface plasmons.

As for the mode with $E_z=0$, Eq.(\ref{Hequation}) is reduced to
\begin{equation}
\gamma E_y=\mathrm{i}\omega\mu_0 H_x\ , \quad
-\gamma E_x=\mathrm{i}\omega\mu_0 H_y\ .
\end{equation}
Following the calculations as shown above,
one can derive the expression of the dispersion relation of the surface plasmons as
\begin{equation}
\begin{split}
&\big(\gamma^{(1)}+\gamma^{(2)}-\mathrm{i}\omega\mu_0\sigma_{xx}\big)
\big(\gamma^{(1)}+\gamma^{(2)}-\mathrm{i}\omega\mu_0\sigma_{yy}\big)
\\&+(\omega\mu_0)^2\sigma_{xy}\sigma_{yx}=0\ .
\end{split}
\end{equation}
Considering the symmetry of the BHZ model, this expression can be further written as
\begin{equation}
\big(\gamma^{(1)}+\gamma^{(2)}-\mathrm{i}\omega\mu_0\sigma_{xx}\big)^2
=(\omega\mu_0\sigma_{xy})^2\
\label{ctensor}
\end{equation}
which can be easily solved. We have found that it is more convenient working with complex angular frequency.
Replacing $\omega$ with $\omega_{\mathrm{R}}-\mathrm{i}\omega_{\mathrm{I}}$, where
$\omega_{\mathrm{R}}$ and $\omega_{\mathrm{I}}$ are respectively the real and imaginary parts,
Eq.(\ref{ctensor}) can be broken down into two branches of dispersion relations written as follows:
\begin{equation}
\begin{split}
&\gamma^{(1)}+\gamma^{(2)}\\&=\frac{[(\mathrm{Re}(\sigma_{xy})-\mathrm{Im}(\sigma_{xx}))^2+
(\mathrm{Im}(\sigma_{xy})+\mathrm{Re}(\sigma_{xx}))^2]\omega_{\mathrm{R}}\mu_0}
{\mathrm{Re}(\sigma_{xy})-\mathrm{Im}(\sigma_{xx})}
\end{split}
\label{set1}
\end{equation}
with
\begin{equation}
\omega_{\mathrm{I}}=\frac{\mathrm{Im}(\sigma_{xy})+\mathrm{Re}(\sigma_{xx})}
{\mathrm{Re}(\sigma_{xy})-\mathrm{Im}(\sigma_{xx})}\omega_{\mathrm{R}}\ ,
\label{set2}
\end{equation}
and
\begin{equation}
\begin{split}
&\gamma^{(1)}+\gamma^{(2)}\\&=-\frac{[(\mathrm{Re}(\sigma_{xy})+\mathrm{Im}(\sigma_{xx}))^2+
(\mathrm{Im}(\sigma_{xy})-\mathrm{Re}(\sigma_{xx}))^2]\omega_{\mathrm{R}}\mu_0}
{\mathrm{Re}(\sigma_{xy})+\mathrm{Im}(\sigma_{xx})}
\end{split}
\label{set3}
\end{equation}
with
\begin{equation}
\omega_{\mathrm{I}}=\frac{\mathrm{Im}(\sigma_{xy})-\mathrm{Re}(\sigma_{xx})}
{\mathrm{Re}(\sigma_{xy})+\mathrm{Im}(\sigma_{xx})}\omega_{\mathrm{R}}\ .
\label{set4}
\end{equation}
These two solutions are related by time-reversal symmetry. Since $\gamma^{(1)}+\gamma^{(2)}>0$,
$\mathrm{Re}(\sigma_{xy})-\mathrm{Im}(\sigma_{xx})>0$ and
$\mathrm{Re}(\sigma_{xy})+\mathrm{Im}(\sigma_{xx})<0$
must be satisfied in Eqs.(\ref{set1})
and (\ref{set3}), respectively. Also,
$\omega_{\mathrm{I}}$ must be positive.

\section{Results}\label{section4}
The Fermi energy locates within the gap, thus only interband transition of electrons needs to be considered during the calculations of the conductivities. Due to the absence of the intraband transitions, such optically conductive surface resembles a dielectric thin film rather than a metallic one. This fact can also be known from the imaginary parts of the longitudinal conductivities as shown in Fig.\ref{fig2} which is negative leading to positive effective permittivities. Without parity anomaly, the Hall conductivities with positive and negative Dirac mass terms differ just by a sign, as shown by the solid red and blue solid curves in Fig.\ref{fig2}(a). In the presence of parity anomaly, the Hall conductivities are respectively zero and integer-valued in the topologically trivial and non-trivial situations.

We have searched for the two surface plasmon modes mentioned in Sec.\ref{section3}, and for simplicity the surrounding dielectrics have been assumed to be isotropic with refractive indices $n=3$, which is reasonable and will not affect our main conclusions in this paper. The dispersion relations of the surface plasmon mode with $H_z=0$ are shown in Fig.\ref{fig3}, where we have solved the dispersion relations in the complex plane and plotted the propagation constant $\beta$ as functions of the photon energy. Figs. \ref{fig3}(a) and \ref{fig3}(b) correspond to $mv^{2}=0.05$ and $mv^{2}=-0.05$, respectively. The real and imaginary parts are plotted in blue and red, respectively. Propagation constants correspond to the case with/without parity anomaly are plotted with
dashed/solid curves. Insets are the enlarged plots of the region below the transition threshold. Since the conductivities given by the BHZ model are not Drude-type, the dispersion curves as shown in Fig.\ref{fig3} are either straight lines corresponding to the light line in surrounding dielectrics or curves with relatively large imaginary parts. Straight lines indicate the absence of any surface-confined electromagnetic modes, and $\mathrm{Im}(\beta)\gg\mathrm{Re}(\beta)$ simply means large energy dissipations. For $mv^{2}=0.05$, the parity anomaly seems to lower the curves of the propagation constants, as shown in Fig. \ref{fig3}(a); while for $mv^{2}=-0.05$ the propagation constants are increased, as shown in Fig. \ref{fig3}(b). It is clear that the parity anomaly in a two-dimensional Chern insulator can seriously modify the dispersion curves of this surface plasmon mode.

As for the surface plasmon mode with $E_z=0$, two branches of dispersion relations have been found by numerically solving Eqs.(\ref{set1}) and (\ref{set3}). The dominators of these two equations are respectively plotted as functions of the photon energy in Figs.\ref{fig4}(a) and \ref{fig4}(b). From a mathematical point of view, Eqs.(\ref{set1}) and (\ref{set3}) respectively require $\mathrm{Re}(\sigma_{xy})-\mathrm{Im}(\sigma_{xx})>0$
and $\mathrm{Re}(\sigma_{xy})+\mathrm{Im}(\sigma_{xx})<0$. Without parity anomaly, as shown by the solid curves in Figs. \ref{fig4}(a)and \ref{fig4}(b), $mv^{2}=0.05$ and $mv^{2}=-0.05$ actually give the same dispersion curves, i.e. they are degenerate.
However, with parity anomaly, such degeneracy is lifted, as shown by the dashed blue and red curves. For the branch corresponding to Eq.(\ref{set1}), both $mv^{2}=0.05$ and $mv^{2}=-0.05$ can lead to physical solutions since $\mathrm{Re}(\sigma_{xy})-\mathrm{Im}(\sigma_{xx})>0$; while for the branch corresponding to Eq.(\ref{set3}), only $mv^{2}=-0.05$ can lead to meaningful results where $\mathrm{Re}(\sigma_{xy})+\mathrm{Im}(\sigma_{xx})<0$.

The right hand sides of Eqs.(\ref{set1}) and (\ref{set3}) are solely determined by the photon energy which are plotted in Figs.\ref{fig4}(c) and \ref{fig4}(d). Without parity anomaly, $\gamma^{(1)}+\gamma^{(2)}$ of the two branches corresponding to Eq.(\ref{set1}) ($mv^{2}=0.05$) and Eq.(\ref{set3}) ($mv^{2}=-0.05$) are the same, as indicated by the fact that the solid blue curve in Fig.\ref{fig4}(c) and solid red curve in Fig.\ref{fig4}(d) coincide. With parity anomaly, as shown by the dashed curves in Figs.\ref{fig4}(c) and \ref{fig4}(d), the branch corresponding to Eq.(\ref{set1}) has two solutions, where dashed blue and red curves in Fig.\ref{fig4}(c) respectively denote the situations with $mv^{2}=0.05$ and $mv^{2}=-0.05$; while the branch corresponding to Eq.(\ref{set3}) has only one solution as indicated by the red dashed curve in Fig.\ref{fig4}(d) denoting the situation with $mv^{2}=-0.05$.

These surface plasmons are weakly guided based on the observation that the dispersion relations are quite close to the light line in the surrounding dielectric material. We have solved the mode effective indices defined as $\beta/k_0$ as functions of the photon energy, which are plotted in Figs.\ref{fig5}(a) and \ref{fig5}(b). Again, without parity anomaly, two branches of this surface plasmon mode are degenerate and the solid blue and red curves in Figs. \ref{fig5}(a) and \ref{fig5}(b) are identical; such degeneracy can be lifted by introducing the parity anomaly term in the Hamiltonian. The effective index of the surface mode should slightly larger than the refractive index of the surrounding dielectric material. We further plot $\omega_i/\omega_r$ as functions of the photon energy in Figs.\ref{fig5}(c) and \ref{fig5}(d). $\omega_i/\omega_r>0$ must be satisfied since the energy must be damped during propagation. Large ratios mean that these surface plasmons possess significant losses which can be ascribed to relatively small conductivities given by the BHZ model.

\section{Conclusion}\label{section5}
In this paper, we have investigated the relations between the parity anomaly in a two-dimensional Chern insulator and the dispersion relations of the surface plasmons. Given by the symmetry of the model we have considered, two surface plasmon modes have been found. Each mode contains two branches of dispersion relations which are degenerate with regard to the sign of the Dirac mass term in the absence of the parity anomaly. Introducing the parity anomaly term into the Hamiltonian will lift this degeneracy and significantly modify the dispersions of the surface plasmons. In the presence of the parity anomaly, the band topology of the bulk states results in integer-valued Hall conductivity. Despite the fact that the Hall conductivity is shifted about $e^2/2h$, it can cause significant changes and even leads to the occurrence of an extra branch of surface plasmons. Our findings have revealed the connections between the parity of two dimensional materials and the dispersion relations of their surface plasmons, which might become valuable in, for example, detection of the parity anomaly via plasmonic responses.

\acknowledgments
This work was supported by National Natural Science Foundation of China (Grant Nos.11804070, 61805062, 11975088).

\end{document}